# Pseudomagnetic fields and strain engineering: graphene on GaN nanowires

Jakub Kierdaszuk[1*], Paweł Dąbrowski[2], Maciej Rogala[2], Paweł Krukowski[2], Aleksandra Przewłoka[3,4,5], Aleksandra Krajewska[3,4], Wawrzyniec Kaszub[3], Marta Sobanska[6], Zbigniew R. Zytkiewicz[6], Vitaly Z. Zubialevich[7], Paweł J. Kowalczyk[2], Andrzej Wysmołek[1], Johannes Binder[1], Aneta Drabińska[1]

[1]Faculty of Physics, University of Warsaw, Poland

[2]Faculty of Physics and Applied Informatics, University of Lodz, Poland

[3]Łukasiewicz Research Network - Institute of Microelectronics and Photonics, Centre for Electronic Materials Technology, Warsaw, Poland

[4]Center for Terahertz Research and Applications (CENTERA), Warsaw, Poland

[5]Institute of Optoelectronics, Military University of Technology, Warsaw, Poland

[6]Institute of Physics, Polish Academy of Sciences, Warsaw, Poland

[7]Tyndall National Institute, University College Cork, Cork, Ireland

*Corresponding author. E-mail: jakub.kierdaszuk@fuw.edu.pl

Abstract

Gallium nitride nanowire and nanorod substrates with different morphology are prospective platforms allowing to control the local strain distribution in graphene films top of them, resulting in an induction of pseudomagnetic fields. Atomic force microscopy measurements performed in a HybriD mode complemented by scanning electron microscopy allow for a detailed visualization of the strain distribution on graphene surface. Graphene in direct contact with supporting regions is tensile strained, while graphene located in-between is characterized by lower strain. Characteristic tensile strained wrinkles also appear in the areas between the supporting regions. A detailed analysis of the strain distribution shows positive correlation between strain gradient and distances between borders of supporting regions. These results are confirmed by Raman spectroscopy by analysis the D' band intensity, which is affected by an enhancement of intravalley scattering. Furthermore, scanning tunneling spectroscopy shows a local modification of the density of states near the graphene wrinkle and weak localization measurements indicate the enhancement of pseudomagnetic field-induced scattering. Therefore, we show that nanowire and nanorod substrates provide strain engineering and induction of pseudomagnetic fields in graphene. The control of graphene morphology by a modification of distances between supporting regions is promising for both further fundamental research and the exploration of innovative ways to fabricate pseudomagnetic field-based devices like sensors or filters.

Introduction

The precise control of the strain distribution in graphene membranes opens up new pathways for the modification the electronic structure of graphene, bandgap opening, and the creation of pseudomagnetic field.[1,2,3,4,5] A non-uniform elongation of the graphene lattice constant leads to the formation of an effective gauge potential, so that a local pseudomagnetic field up to hundreds of tesla can be induced.[6,7] The pseudomagnetic field at two nonequivalent graphene K and K' Dirac cones has opposite sign, therefore valley filter devices have been recently investigated theoretically and experimentally.[8,9,10,11] The probability of intravalley scattering in graphene can be enhanced by an interaction with a pseudomagnetic field as reported by measurements of a Raman D' band intensity enhancement, suppression of weak localization (WL) signal and a decreasing of scattering lengths.[12,13] This large field is also important for studies of pseudo-Landau quantization, topological states, and the fabrication of next-generation sensors.[1,7,14] Several methods of non-uniform graphene deformation like graphene bending, crumbling, or folding were proposed.[12,15,16] Deposition of graphene on nanoparticles or nanopyramids enables the precise control of graphene roughness, and strain gradient.[17,18,19] Inhomogeneous applied strain results in wrinkles formation which can be traced by atomic force microscopy (AFM) measurements and analyzed by numerical simulations of strain distribution.[18,19] Theoretical calculations can also predict the value of pseudomagnetic field induced in locally strained graphene in the function of the absolute value of strain, its orientation, and graphene morphology.[15,19] Standard confocal Raman measurements of pseudo-Landau states in graphene require areas with a diameter of hundreds of nanometers characterized by a homogenous distribution of pseudomagnetic field which is difficult to achieve in typical experiments.[15] Therefore, scanning tunneling spectroscopy was proposed for tracing the local heterogeneity of electronic structure.[7,20] Nevertheless, pseudo-Landau quantization or pseudomagnetic field-induced topological states measured by STM were previously reported only in samples containing graphene wrinkles of bubbles of height lower than 30 nm.

Recent studies show that array of graphene wrinkles can be obtained on lithographically etched nanorods substrate or epitaxially grown nanowires.[21,22] Gallium nitride (GaN) nanorods or nanowires covered with graphene are promising not only for the fabrication of efficient nano-light-emitting diodes or UV sensors but also as a platform for surface-enhanced Raman scattering experiments.[23,24,25,26,27,28] GaN structures can be prepared by top-down (lithography + dry etch) or bottom-up (epitaxial growth) approaches under different conditions providing control over the rod diameters, their interdistances and height variations.[21,29,30,31] These parameters are crucial for the modification of graphene wrinkle morphology and consequently the spatial distribution of pseudomagnetic fields.[21,32,33]

A theoretical model of the pseudomagnetic field dependence on the graphene bend radius shows a negative correlation.[34] The local modulation of strain in nanometer-scale graphene wrinkles induces a higher pseudomagnetic field than in graphene folds characterized by bending radius larger than tens of nanometers. Thus, a precise analysis of graphene wrinkles formation on nanopillars and graphene stiffness with a high spatial resolution is necessary to identify local modulation of strain and the possibility of induction of large pseudomagnetic field. AFM measurements of the elastic membrane are, however, complicated due to an interaction between the AFM tip and the elastic suspended graphene as well as gas adsorbents on a rough graphene surface. The local modulation of strain also affects graphene stiffness which is not easily visible in topography measurements performed using tapping mode AFM. Moreover, the large unevenness of the nanorods substrate makes precise and stable scanning tunneling spectroscopy (STS) measurements of pseudo-Landau quantization with nanoscale resolution nontrivial.[17] Nanorods could also cause local cracks in monolayer graphene after its deposition which further complicates the analysis of WL signal.[28]

In our work, we present a study of pseudomagnetic field generation in two-layer graphene suspended on epitaxially grown GaN nanowires (NWs) and plasma etched GaN nanorods (NRs). These results are followed by a detailed analysis of wrinkles formation and strain distribution in graphene. We mainly focus on samples characterized by different interdistances between NWs/NRs on top of which graphene is supported. The impact of other parameters: differences in NWs and NRs heights, diameter, distances between NWs/NRs borders and their arrangement regularity on the graphene properties are also analyzed. Two-layer graphene was used to obtain a large area of the high-quality transferred graphene on the corrugated substrate. Because contactless WL measurements require a larger area of graphene sheet area, tri-layer graphene was also used and compared with two-layer samples. The first part of our work provides a detailed study of the sample topography. Scanning electron microscopy (SEM) was used to measure all NW/NR samples (bare and covered with graphene). The SEM results were complemented by AFM HybriD mode measurements which allow measuring the sample topography with a high resolution and the mechanical interaction between graphene and tip simultaneously.[35] This approach provides valuable information about sample topography, stiffness, and deformation as well as their spatial distribution and correlation with NWs and NRs morphology. Stiffness and deformation are correlated with the value of graphene strain.

In the second part, manifestations of the occurrence of pseudomagnetic field are investigated. Raman spectroscopy was used to study the intensity of graphene D' Raman band for tracing a possible fingerprint of the pseudomagnetic field-induced enhancement of intravalley scattering in graphene. STS measurements near graphene wrinkle were performed to study changes in the local density of states (LDOS) of deformed two-layered graphene. Raman and STS studies were supported also by analysis of weak localization effect measured by contactless microwave-induced electron transport (MIET).[27,36] This approach enables to make electron transport measurements of fragile samples without the fabrication of electric contacts. Scattering on defects can change the graphene pseudo and isospin resulting in a reduction of backscattering and consequently sample resistivity.[37] Analysis of the weak localization signal suppression and the reduction of the characteristic scattering lengths caused by an interaction with pseudomagnetic field was performed. We demonstrate that NWs and NRs interdistances and diameters strongly affect the wrinkle morphology, the presence of multiple wrinkles between NRs and a local gradient of graphene strain. Moreover, those wrinkles and graphene strain gradients could be responsible for the induction of local and strong pseudomagnetic fields in graphene on NWs and NRs which affect the properties of partially supported graphene.

Sample's details

Two-layer graphene was transferred from copper onto different substrates of GaN NWs and NRs by a polymer frame method.[38] Three samples of graphene on NWs with interdistances (distances between the centers of the supporting regions) between nearest NWs supporting graphene estimated as 93, 145, and 285 nm were named as NWs90, NWs150, and NWs290, respectively.[30] All samples' details obtained from analysis of SEM images are included in Table 1. In the NWs90 sample, NWs have a nearly equal height of 900 nm. NWs form small clusters with average distances of 200 nm between them and density 25 clusters/$\mu m^2$. In NWs150 and NWs290, NWs have non-equal height and only small percentage of the highest NWs are in contact in graphene. Therefore, discussed sample parameters are estimated only for these high NWs. For the NWs150 sample, 19 % of all NWs are 100 nm higher than the rest and form small groups with at a distance less than 100 nm. For the third sample NWs290, the NWs have two different heights. Approximately 10% of NWs are about 1.5 μm height while the rest are about 1 μm height. The distances between rarely arranged high NWs vary from 0.25 to 1 μm. The diameters of all NWs are similar in the range 34-45 nm.

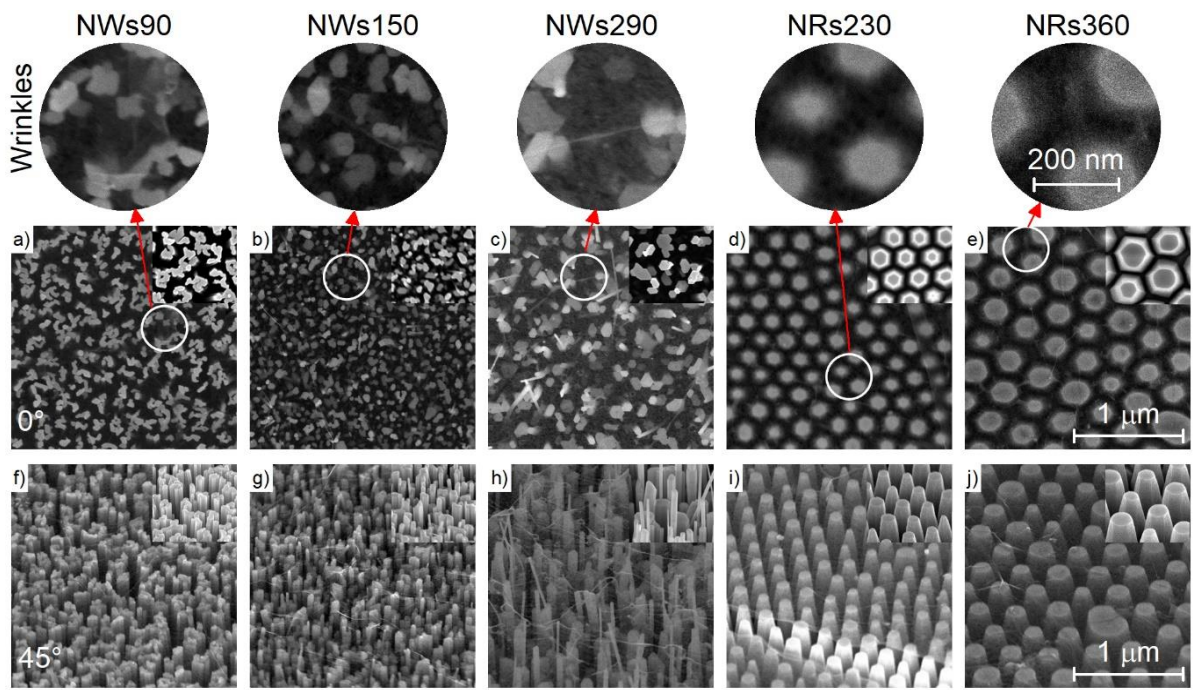

Figure 1. SEM image of graphene on NWs with 90 nm (a, f), 150 nm (b, g), and 290 nm (c, h) interdistances and NRs with interdistances around 230 nm (d, i) and 360 nm (e, j); performed at planar view (a-e) and 45° tilt (f-j). Insets present NWs and NRs substrates without graphene. Circles depict graphene wrinkles visualized on enlarged images on the top.

The other two samples were prepared using substrates consisting of NRs with equal height (800 nm) (Fig. 1).[39] For the first sample (NRs230), GaN NRs have an average diameter of their top $c$-plane facet of 132 nm and interdistances of 232 nm (Table 1). For the second sample (NRs360), the diameter and interdistances are about 1.5 times larger. Therefore, all NWs and NRs samples differ in interdistances between supporting regions. NRs have also larger diameter than NWs. Therefore, the area of graphene suspended between the corresponding supporting regions should be characterized by an additional parameter: a distance between the borders of supporting NWs/NRs (b-b). More details about NWs/NRs growth and preparation of two-layered graphene are included in Methods section.

Table 1. Comparison of NWs and NRs parameters: diameters (d), interdistances (id, distances between the middle of the nearest NWs), number of NWs/NRs per μm$^2$ (wd), border-to-border (b-b) distances (distances between borders of closest NWs/NRs). For NWs150 and NWs290, parameters are presented only for the NWs in contact with graphene. The last two columns present a comparison of the percent of the area of graphene supported by the substrate (AGS), obtained by analysis of SEM images (Fig. 1) and AFM HybriD mode data (Fig. 2).

| Samples | d (nm) | id (nm) | wd (/μm$^2$) | b-b (nm) | AGS SEM (%) | AGS AFM (%) |
|---|---|---|---|---|---|---|
| NWs90 | 45 | 93 | 150 | 48 | 37 | 19 |
| NWs150 | 45 | 145 | 75 | 100 | 11 | 11 |
| NWs290 | 34 | 285 | 10 | 251 | 5 | 2 |
| NRs230 | 132 | 232 | 20 | 100 | 29 | 26 |
| NRs360 | 195 | 360 | 9 | 165 | 27 | 27 |

Results

Sample topography

Figure 1 shows an SEM image of the morphology of graphene deposited on NWs and NRs. Compared to pristine NWs and NRs, graphene looks like a grey sheet draped over supporting NWs and NRs like a tent (Fig. 1a-j). This is visualized clearly in Figure 1i which shows NRs230 with areas covered and uncovered by graphene. Different brightness of these areas enables the recognition of the border of the graphene sheet. In general, SEM images performed at lower magnification and tapping mode AFM confirms a high uniformity of obtained samples occasionally spread out with graphene cracks as presented in Supplementary Materials (Fig. S1 and Fig. S2).

Graphene on NWs is in contact only with the highest NWs, even in the case of NWs90, which is characterized by only low variations of NWs height and small distances between nearest NWs (Fig. 1a, f). Graphene wrinkles connect the nearest supporting regions which is better shown on the image collected from a tilted sample (Fig. 1f). Wrinkles are clearly visible in NWs150 and NWs290 samples with larger distances between nearest NWs supporting graphene (Fig 1b, c, g, h). This result is consistent with a previous report for graphene on silicon nanopillars with different interdistances.[32] Variations in graphene height also directly positively correlate with a decrease of AGS values listed in Table 1. Graphene deposited on the well-ordered hexagonal structure of GaN NRs is smoother than on NWs (Fig. 1d, e, i, j) and with well-defined wrinkles between adjacent NRs. This effect is similar to that reported in graphene deposited on nanoparticles arranged in a hexagonal lattice.[18] Some wrinkles are double (indicated by circles in Fig. 1d, e) which is not observed for NWs samples. Probably, a significantly larger diameter of supporting regions is necessary for the creation of multiple graphene wrinkles. Thus, SEM shows a positive correlation between graphene roughness and the presence of wrinkles which depends on NWs/NRs substrate morphology. The substrates with regular arrays of NRs have also more regularly defined wrinkles. Moreover, it is observed that graphene is wrapped over the highest NWs, therefore distinguishing NW in contact with the graphene of those distant from it in some cases is nontrivial. Although initial conclusions about the overall morphology can be drawn from SEM measurements, the resolution and spatial contrast are not high enough to satisfactory visualize the morphology of wrinkles and convey more information about local strain distribution.

HybriD mode AFM

HybriD mode AFM was applied to overcome SEM limitations and better visualize the morphology of graphene wrinkles.[35] This approach is important for further discussion of the occurrence of pseudomagnetic field in graphene wrinkles.[18] With this technique, simultaneous acquisition of sample topography, stiffness and deformation are possible. In our case, the graphene membrane is wrapped over NWs/NRs and depending on their interdistance the value of local strain in graphene varies between suspended and supported regions which should modify the susceptibility of graphene to elastic deformation. Graphene stiffness and deformation calculated from the measured force-distance curves differ between strained and relaxed areas.[40] The area of highly strained graphene is characterized by high stiffness and low deformation while in fully suspended graphene these parameters should reach opposite values. Standard models are used for the analysis of distance-force curves based mainly on the assumption of local interaction between the AFM tip and the surface.[41,42] However, in our case, suspended graphene is elastically deformed by the supporting areas (NWs/NRs top facets) spaced at different distances from the local point of contact. Therefore, the measured stiffness and deformation visualizes the relative differences in local graphene strain distribution and differences between samples, but their recalculation into absolute values is non-trivial. For this reason, the units of measured stiffness and deformation will be denoted as arbitrary (a. u.).

An analysis of the AFM topography data measured for all investigated samples reveals substantial height variations of graphene deposited on NWs (Fig. 2a-e). Deviation of graphene height is correlated with the increase in average interdistances between supporting regions. Graphene wrinkles are hard to recognize. The topography of the NRs samples (Fig. 2d, e) is significantly smoother, with small variations of graphene height.

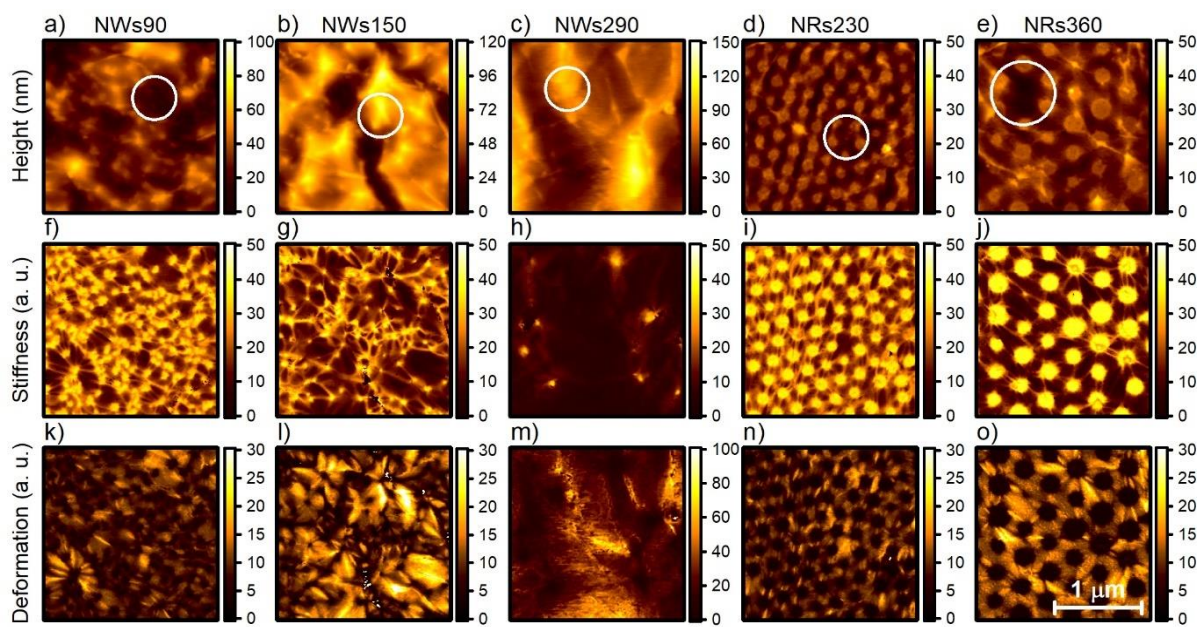

Figure 2. Results of HybriD mode AFM measurements of topography (a-e), stiffness (f-j), and deformation (k-o) of samples: NWs90 (a, f, k), NWs150 (b, g, l), NWs290 (c, h, m), NRs230 (d, i, n), NWs360 (e, j, o). The areas depicted by white circles are discussed in the text.

Maps of sample stiffness (Fig. 2f-j) visualize a strikingly different picture including the identification of areas characterized by high stiffness. These areas are related to graphene-supported on NWs/NRs linked by wrinkles of graphene with medium stiffness. There are more NWs and wrinkles visible in stiffness maps as compared to topography images (Fig. 2a-e). A comparison with SEM is presented in Supplementary Materials (Fig. S3). This observation is clear evidence of the presence of wrinkles of tensely strained graphene which connect nearest-neighbor NWs and NRs. This behavior is similar to the morphology of graphene deposited on the surface of the nanoparticles, where supporting regions are also connected with wrinkles.[18] The HybriD mode technique enables the identification of these stiffer areas, visualizes their position with higher resolution than SEM (Fig. 1d, e) and allows to conduct the more accurate analysis. One example is marked by the white circle for the NWs90 (Fig. 2a) for which the stiffness map (Fig. 2f) shows the presence of more NWs and wrinkles than the topography map. In the NWs150 sample, HybriD mode also enables a better estimation of the number of NWs in direct contact with graphene (see the white circle in the topography image of NWs150, Fig. 2b). Moreover, some NWs are closer to each other and are connected by less visible wrinkles than the rest. This suggests that wrinkles are more easily formed in graphene suspended between NWs clusters than in graphene deposited on a cluster of NWs where interdistances are smaller. Analysis of the NWs290 sample underlines, that stiffness analysis enables to recognize graphene suspected of being supported by NWs (white circle in Fig. 2c) from being truly supported by NWs (Fig. 2h). However, wrinkles in NWs290 samples are better visible in SEM than in HybriD mode AFM. This is probably related to a higher variation of height and stiffness in sparsely supported graphene which makes isolation of wrinkles difficult. NRs diameters are 3-4 times larger compared to those of NWs, another difference is that an absolute majority of NRs are also characterized by the same heights (except those that do not have $c$-plane facets). Although the position of a single NRs is well defined, an analysis of stiffness maps (Fig. 2i, j) allows designating the position of some NRs with lower diameter and height which are not observable in topography images (e.g. in the white circled area in Fig. 2e). Stiffness maps visualize graphene wrinkles with high accuracy, some of them are double or triple (Fig. 2j). Because of the higher uniformity of NRs compared to NWs, the arrangement of wrinkles is also relatively well ordered in contras to those in NWs samples.

The graphene deformation maps correspond well with the stiffness results. NWs, NRs, and graphene wrinkles are characterized by low deformation factors, while the highest deformation occurs in areas between graphene wrinkles and NWs/NRs. This is related to the greater susceptibility of locally unsupported graphene to deflection during contact with the AFM probe. For the NWs90 sample, graphene reaches higher deformation only near the highest NWs (Fig. 2k). Wrapping of graphene over these tall NWs detaches graphene from the closely located lower NWs. The local increase of interdistances between supported regions is responsible for the formation of wrinkles composed of higher tensile strain, and for the formation of larger areas of relaxed suspended graphene. Similar behavior occurs for the NWs150 sample (Fig. 2l); however, this sample is characterized by higher variations in the interdistances lengths. Therefore, the variation of graphene deformation between adjacent suspended graphene areas is also larger. This result confirms that a local increase in the density of supporting regions reduces graphene deformation, indicated by the higher average value of graphene strain. Graphene deformation for the NWs290 sample (Fig. 2m) is the highest of all samples which is related to the sparse arrangement of supported NWs. The deformation distribution in NRs samples is more uniform than in NWs samples. In suspended graphene areas the average value of deformation in NRs360 is higher than in NRs230. Moreover, its value increases in the presence of vacancies in the NRs array, as shown in Fig. 2d where the white circle depicts a dislocation in the NRs230 sample. This vacancy results in higher NRs interdistances than between uniformly distributed NRs. Therefore, the deformation factor of graphene for this vacancy is larger than in the nearest area between surrounding NRs. Graphene wrinkles over the vacancy are less significant, but still present (Fig. 2i, n). The white circle (Fig. 2d) depicts another interesting place of the NRs360 sample where a lower and thinner nanorod is present. Four wrinkles connected closely at the center of the area are visible in the stiffness and deformation maps (Fig. 2j, o).

The HybriD mode data on sample deformation was also converted into histograms which are presented in Fig. 3. In each histogram, there were identified three or four different components originating from graphene: supported on NWs/NRs, graphene very close to NRs borders, and two different components for suspended graphene areas. They were denoted as NW/NR, borders, SG1 and SG2, respectively. The SG1 component is characteristic for graphene located between closely situated NWs/NRs while the SG2 component characterizes graphene located far away from supported regions. The SG2 component is often present over vacancies in NRs array or in the vicinity of the highest NWs. A complementary and consistent analysis justifying this approach was included in Supplementary Materials (Fig. S4). The NW/NR deformation component is characterized by a sharp peak of the low deformation factor. Graphene deposited directly on GaN adheres to its surface, therefore it is strained compressively, and the deformation should be equal to zero. However, the local roughness of NW/NR surface allows to deform graphene slightly. The borders component of the histograms is present only for graphene deposited on NRs and has slightly higher deformation than NR component. Graphene directly surrounding the supporting area is under additional tensile strain caused by the adhesion of graphene to the side surface of NR. The small diameter of NWs compared to NRs leads probably to a decrease of the area of this component which disables its identification in NWs samples. The SG1 and SG2 components correspond to the suspended graphene characterized by high deformation. The results of all fitted component (its positions and percentage contribution) are summarized in Table 2.

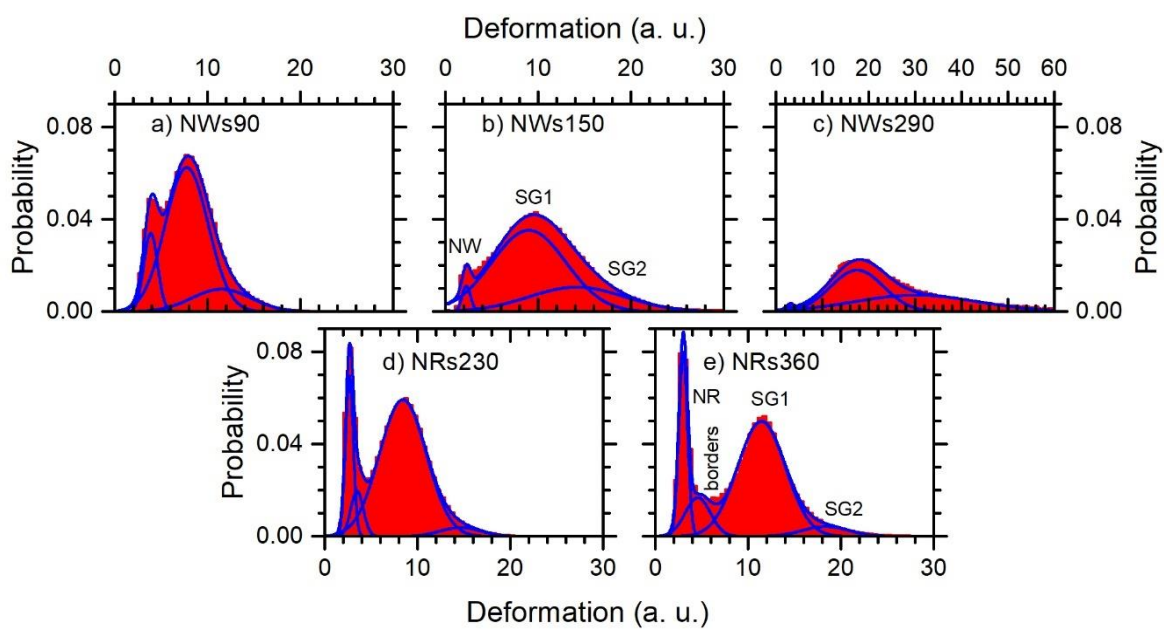

Figure 3. Histograms of graphene deformation obtained from HybriD mode AFM analysis for the measured samples: a) NWs90, b) NWs150, c) NWs290, d) NRs230, e) NRs360. Captions

at point b) and e) depicted components analyzed in text. The horizontal scale in NWs290 differs from the rest samples due to the higher width of the histogram.

Table 2, Maximum position and area (as percent of total histogram area) of each component in the deformation distribution for all the samples. The components are indicated as: NW/NR – graphene deposited directly on NWs and NRs, Border – graphene located directly near NR edge, SG1 and SG2 – the suspended graphene. Additionally, the percentage ratio of SG2/SG1 is included.

| | Maximum position (a.u.) | | | | Component area (%) | | | | Area ratio |
|---|---|---|---|---|---|---|---|---|---|
| Sample | NW/NR | Borders | SG1 | SG2 | NW/NR | Borders | SG1 | SG2 | SG2/SG1 |
| NWs90 | 4 | - | 8 | 12 | 13 | - | 72 | 15 | 0.21 |
| NWs150 | 2 | - | 9 | 15 | 3 | - | 70 | 27 | 0.39 |
| NWs290 | 3 | - | 18 | 30 | 1 | - | 51 | 48 | 0.94 |
| NRs230 | 3 | 4 | 8 | 15 | 15 | 7 | 74 | 4 | 0.05 |
| NRs360 | 3 | 5 | 11 | 18 | 19 | 12 | 64 | 5 | 0.08 |

An analysis of the shift of the positions of the maxima and the percentage of points located within these components allow establishing the impact of NWs/NRs substrate parameters on the graphene strain distribution. The main parameters are the border-to-border (b-b) distances between closest NWs/NRs, and the percentage of the area of graphene supported (AGS) presented in Table 1. A positive correlation between the last mentioned and the contribution of NW/NR component area is observed. Some discrepancies between these two values result from overlapping of different components. This effect is significantly visible for the NWs150 sample where the SG1 component is wide and overlaps with the NW component (Fig. 3b). In the NRs360 sample, a larger diameter of NRs is responsible for the presence of about 1.7 times larger component area of borders points compared to the NRs230 sample.

Arbitrary positions of SG1 and SG2 maxima are positively correlated with b-b in all samples which is the most significant for the samples with the highest b-b. Moreover, the fraction of points in the SG1 component in all samples is negatively correlated with the b-b parameter; however, the percentage of points in SG2 component is correlated with the b-b parameter only for NWs samples. In both NRs samples, the percentage of points in the SG2 component is dozen times lower than SG1 in both NRs samples. These observations allow drawing several conclusions. First, an increase of the average b-b distance is responsible for the larger strain gradient in suspended graphene located directly between supporting regions (related to SG1component). An increase of b-b distance between NWs also increases strain gradient in suspended graphene located in areas where distances between supporting regions are higher (related to SG2 component). Contrary to NWs, lithographically etched NRs samples are characterized by more uniform morphology and almost no variations in height. Graphene characterized by higher deformation is distributed only at sparsely distributed vacancies in NRs arrangement. The presence of multiple graphene wrinkles also increases the uniformity of deformation which is responsible for the decrease of the width of the SG1 and SG2 components. Therefore, all samples are characterized by an additional parameter: the area ratio of SG2/SG1. A low value of this parameter in NRs samples is a fingerprint of a small area of gradient between graphene suspended directly between NRs and graphene on areas characterized by higher b-b than average. Higher variations of b-b in NWs samples are also correlated with an increase of gradient between the rate of suspended areas and the SG2/SG1 ratio.

Thus, the statistical analysis of the HybriD mode AFM map of graphene deformation links the distances between borders of supporting regions and their diameters with the variations of graphene strain and its spatial distribution. NWs and NRs substrate morphology effectively creates graphene wrinkles. The repeatability of wrinkles distribution depends on the degree of order of supporting regions. Supporting regions with large diameters can generate multiple wrinkles which locally reduces the characteristic size of graphene strain variations. However, the increase of the graphene strain gradient on graphene wrinkles is mainly caused by an increase of distances between NWs/NRs borders (b-b). Moreover, the ratio of suspended graphene areas characterized by higher and lower measured deformation SG2/SG1 varies between samples. In highly oriented NRs samples this value is low. SG2/SG1 significantly increases in unevenly distributed NWs as a function of NWs b-b distance. An area in the vicinity of different NWs enables the collection of data with a wide range of strain gradients, however with a poor repeatability. In NRs samples, wrinkle distribution is more uniform and varies only slightly. These substrates offer the possibility for a local modification of the wrinkle geometry by creating a NR vacancy. In summary, both, NWs and NRs substrates are promising for the creation and control of local strain in graphene which opens novel ways to control the occurrence of pseudomagnetic field which is examined in the second part of the paper.

Raman results

The analysis of the graphene band positions and intensities is performed to estimate the strain behavior in Raman maps and for further investigation of manifestations of the presence of pseudomagnetic field. The Raman shift of the 2D and G bands varies between representative spectra measured on different spots on the sample which suggested the presence of strain variations.[43] Analysis of these shifts is consistent with results of HybriD mode AFM and is discussed in details in Supplementary Materials (Fig. S5).

Raman spectra of graphene on NWs and NRs have a low intensity of defect-induced D and D' bands compared to the intensity of the G band which suggests a low density of defects (Fig. 4a). A detailed analysis of Raman band intensities is applied to trace the presence of pseudomagnetic field-induced enhancement of the intensity of the D' band. Lattice defects in graphene contribute to both: intervalley and intravalley scattering, which are responsible for the creation of double resonant D (~1345 cm$^{-1}$) and D' (~1620 cm$^{-1}$) bands.[44] Intervalley scattering requires a large momentum change, while small momenta are important for intravalley scattering. The creation of the D band is caused by scattering on the iTO phonon and a defect and occurs even if momentum change is large involving two Dirac cones.[45] The D' band appears after the resonant scattering on an electron and /or hole assisted by both, the iLO phonon and elastic defect scattering with small momentum change. Therefore, the creation of a D' band occurs in a single valley so spin flip is essential for backscattering.[45] The lack of this second phenomenon decreases the value of the D' band intensity and affects the experimental value of D' band intensity.

Studies of various types of graphene defects showed that the intensity ratio of the D' band to D band ($R_{DD'}$) varies (depending on the type of lattice defect) from 0.08 for $sp^3$ defects, through 0.14 for single vacancies, 0.29 for grain boundaries, to 0.77 for on-site defects.[46] Impurities of alkali metals can also yield a high ratio of $R_{DD'}$.[45] However, these Coulomb defects cannot affect the carrier phase which suppresses the backscattering and D' intensity below the detection limit. However, strain-induced pseudomagnetic fields can induce a carrier isospin flip which increases the backscattering probability and enhances the D' band intensity. Therefore, the intensity ratio of the D' to D band is significantly higher than for on-site defects and can exceed the value of 0.77. The presence of a D' band enhancement can be taken as a hint of the presence of strong pseudomagnetic field in partially supported graphene.

Two types of Raman spectra measured as part of the Raman map for sample NRs230 are presented in Fig. 4a. These spectra are characterized by similar intensity and positions of D, G, and 2D bands. However, the $R_{D'D}$ ratio enhancement suggests the presence of a local pseudomagnetic field induced by strain variations in graphene wrinkles. Since in our case we study two-layer multigrain graphene with random misalignments, it is worth noting that in two-layer graphene with a lattice angle mismatch between 4° and 6°, the interlayer scattering results in the appearance of an additional graphene R' band with comparable energy and intensity as D' band.[47] However, this case can be distinguished by an analysis of the intensity ratio of the 2D to G band, which also depends on the lattice misalignment of graphene layers.[48] At higher lattice misalignment than 6°, the R' band energy is shifted and its intensity significantly decreases because the laser line comes out of an angle-dependent resonance.[47] Graphene delamination induced by strain or larger misalignments increase the distances between the graphene layers which interact weaker than layers misaligned by a smaller angle than 6 degrees.[34,49] For graphene with large lattice misalignment, the $R_{2DG}$ value is larger than 1, so a peak at 1620 cm$^{-1}$ should not contain any feature of the R' band.[47,48] The $R_{2DG}$ parameter also depends on carrier concentration, however in the regime of the low variations of carrier concentration this factor should affect $R_{2DG}$ less than lattice misalignment.[50] Indeed the G and 2D band analysis did not show large changes in carrier concentration. Therefore, the spectra presented in Fig 4a characterized by $R_{2DG}$ equal to 1.5 should not contain an R' component. To provide further statistical evidence, we present an analysis of the dependence of $R_{D'D}$ on $R_{2DG}$ for all samples (Fig. 4b). The spectra located in section I are influenced by the graphene layer alignment and the presence of an intense R' band. Section II and III consist of spectra with a low intensity of D' bands. The D' intensity in section II could also be affected by the presence of a small intense R' band while spectra located in part III could be related to graphene native defects. Therefore, only part IV contains spectra characterized by an enhanced D' band intensity and a high value of $R_{2DG}$ characteristic of a graphene monolayer or weakly interacting two-layer graphene. These spectra could be significantly affected by an interaction with a pseudomagnetic field induced in strained graphene wrinkles. However, the resolution of Raman spectroscopy is too low to confirm if this effect occurs only on wrinkles and how the $R_{D'D}$ value on wrinkles differs from graphene suspended between NWs or NRs. TERS (tip enhanced Raman scattering) measurements with higher resolution are necessary for a more detailed investigation of this phenomenon. Nevertheless, the effect of interaction with pseudomagnetic field could occur in large areas with a characteristic fingerprint in Raman spectroscopy which is promising for studies using other techniques.

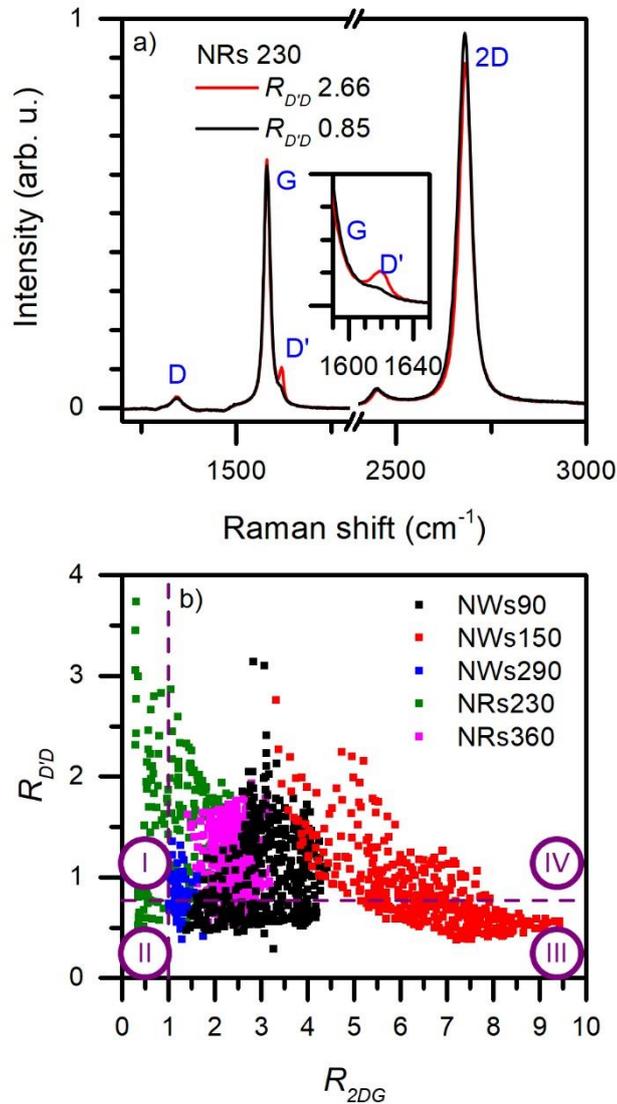

Figure 4. Analysis of pseudomagnetic field enhanced intensity of the D' band versus the ratio of the 2D to G band ($R_{2DG}$) for all samples: a) comparison of Raman spectra with enhanced (red) and non-enhanced (black) D' band for NRs230, the inset shows a zoom on the D' band b) dependence of the intensity ratio of the D' to D band ($R_{D'D}$). The vertical purple dotted line corresponds to the value of $R_{D'D}$ characteristic of on-site defects, while the horizontal line is related to the $R_{2DG}$ limit value characteristic of graphene monolayer or weakly interacting graphene layers.

STS results

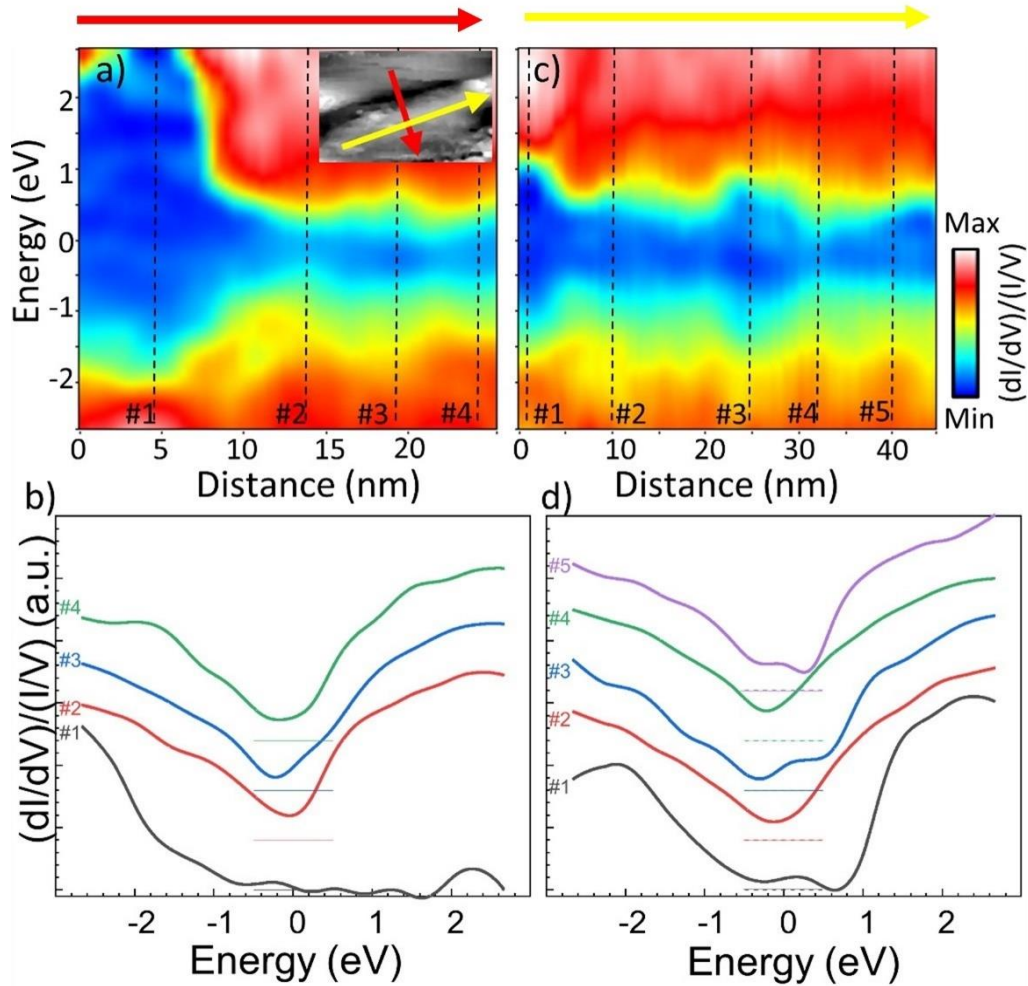

Fig. 6. Normalized tunnelling conductance (nTC) maps recorded along red arrow (a) and yellow arrow (c). Inset in (a) shows 50×25 nm$^2$ STM topography image recorded simultaneously with nTC data. (b) and (d) show individual nTC plots extracted from positions indicated by vertical lines #1-#5 and #1-#4 in maps (a) and (c) respectively.

As has been already shown by AFM (Fig. 2), both types of NWs and NRs substrates induce the formation of graphene wrinkles. The large density of wrinkles in the NWs90 sample (Fig. 2f) makes it promising for STS measurements. This technique was applied for direct access to the electronic structure of graphene on the nanometer scale. In the inset of Fig. 6a, the 50×25 nm$^2$

STM topography image is shown with one of the wrinkles seen in the bottom left (under the yellow arrow). This image was recorded simultaneously with the STS data acquisition. In Fig. 6a, we show normalized tunneling conductance (nTC) map recorded at NWs90 sample in the direction perpendicular to the wrinkle along the line indicated using a red arrow in the inset in Fig. 6a. Inspection of the map reveals clear changes in the local density of the states. These changes are clearly seen for energies above the Fermi level where a clear band gap is present in approx. first 10 nm of the map (see plot #1 in Fig. 6b corresponding to vertical line #1 in Fig. 6a). This indicates the presence of a GaN NW over which part of the map was recorded (beginning of the red arrow in inset in Fig. 6a).[51] This spectrum (plot #1 in Fig. 6b) is characterized by p-doping with a relatively large bandgap starting approx. at 0.5 eV below the Fermi level. The gap seems to be populated by several gap states in agreement to calculation results reported elsewhere.[52]

At approx. 10 nm character of measured spectra changes dramatically (see map in Fig.6a) i.e. bandgap closes and individual plots extracted along lines #2-#4 become graphene-like with characteristic V-shape dependence of nTC on energy (see Fig.6b). Interestingly nTC is changing over the wrinkle which is clearly seen in nTC map shown in Fig. 6c recorded along yellow line shown in the inset in Fig. 6a (see also individual spectra #1-#5 extracted from positions marked by vertical lines in nTC map and shown in Fig. 6d). In particular global minimum which is believed to be correlated to the energy position of the Dirac point (DP) is shifting in energy scale.[53] The DP shift is clearly seen if plots #2, #3, and #4 (Fig.6b) and #1-#5 (Fig.6d) are compared. Mentioned minimum is shifting in range of ±0.3 eV around the Fermi level. This in turn indicates that the DP location is not the same at different points over the wrinkle. Since wrinkles in our experiment are created on graphene supported by NWs separated by dozens of nanometers, we can assume that these wrinkles are freestanding without any support. In consequence, the observed shift of the DP is not related to the charge transfer between graphene and substrate and has to be a manifestation of the local properties of graphene. Interestingly, in some regions along the wrinkle the formation of maximum at the Fermi energy is observed (see Fig. 6c and 6d). Such maxima were previously interpreted as an indication of the signatures of Landau levels formation in consequence of stress applied to graphene.[53] Unfortunately, our data do not allow us to track the formation and shift of other Landau levels reported by Hsu et al.[17] which is probably related to limited resolution in our experiments. Alternatively, the presence of these maxima located at the Fermi level can be the result of flat band formation observed for some magical angles between sheets in two- layered graphene.[54] Despite unknown origin of the maximum at the Fermi level the core observation here is that the electronic structure measured over wrinkle is not homogenous and tend to change.

Weak localization

Contrary to Raman and STS, Microwave Induced Electron Transport (MIET) contactless transport measurements in an ESR spectrometer provide a non-local technique for studies of graphene properties.[36] Due to technical problems in obtaining large macroscopic graphene sheets on substrates with higher distances between NWs, which were necessary for contactless WL measurements, three-layer graphene was prepared using a similar procedure as in the case of two-layer graphene and transferred on NWs and NRs substrates. For the sample with the smallest distances between borders of supporting regions - NWs90, it was possible to obtain also two-layer graphene sample of good morphology. The detailed comparison of WL measurements of those two samples is presented in Supplementary Material (Fig. S6a, b) and shows that the additional layer does not affect the WL signal significantly. Generally, the WL signal depends on intervalley scattering length ($L_i$), long-range scattering length ($L_{lr}$), and coherence scattering length ($L_\varphi$).[37] Elastic scattering lengths ($L_{lr}$, $L_i$) do not depend on temperature, while the inelastic coherence scattering length dependence on temperature can give comprehensive information about the scattering mechanism. Previous works showed that the linear dependence of $L_\varphi^{-2}(T)$ indicates that electron-electron scattering in a diffusive regime, while the square dependence of $L_\varphi^{-2}(T)$ proves the ballistic regime of electron-electron scattering.[55,56] In some works, a zero-offset ($L_{\varphi 0}^{-2}$), was also present in $L_\varphi^{-2}(T)$ dependence and indicated the existence of an additional temperature-independent inelastic scattering.[57,58] The potential cause of this scattering process was previously attributed to spin relaxation due to electron spin-flip scattering which can indicate the existence of a pseudomagnetic field.[12,13]

Intervalley scattering occurs between two Dirac cones and requires isospin flip. The isospin is more robust than the pseudospin so large momentum transfer for scattering between two Dirac cones is necessary. Therefore, sharp point defects or interlayer coupling could enhance this type of scattering.[59,60] Recent works showed that intervalley scattering could also be enhanced by an interaction with a vector potential from magnetic domains or pseudomagnetic field induced by graphene strain.[61,62] In few-layer misoriented graphene, a large in-plane pseudomagnetic field could be induced which also affects the spin texture of graphene and could increase the rate of intervalley scattering.[63] Analysis of crumpled graphene showed increasing intravalley scattering in graphene which was induced by interaction with a pseudomagnetic field.[12] The external vector potential is responsible for a small momentum transfer which enables to flip carrier pseudospin. Therefore, the intravalley scattering rate increases. Similar results were obtained from the analysis of the carrier scattering as a function of strain.[64,65] The discussed process is also responsible for the shortening of the coherence scattering lengths. Therefore, an interaction between charge carrier and pseudomagnetic field changes the graphene pseudospin which affects the scattering rate at one Dirac cone as well as $L_{lr}$ and $L_\varphi$ lengths.

A WL signal was observed for all samples (Fig. 6a) in the range between 5.4 K – 25 K and analyzed by the procedure described in more details in Supplementary Materials and fitted scattering lengths are included in Table 3. MIET measurement in ESR spectrometer give the first derivative of the signal, so signals presented in Fig. 6a were numerically integrated. The directly measured MIET signal is presented in Supplementary Materials in Fig. S6c. When analyzing the amplitude of the MIET signal, one must be careful as it may be affected by variations in graphene area or its placement in the cavity. Moreover, dependence of the signal shape of three scattering lengths is generally non-trivial. However, the significant suppression of the WL signal in NWs290 and NRs230 is observed and suggests the effect of charge carrier interaction with a pseudomagnetic field.

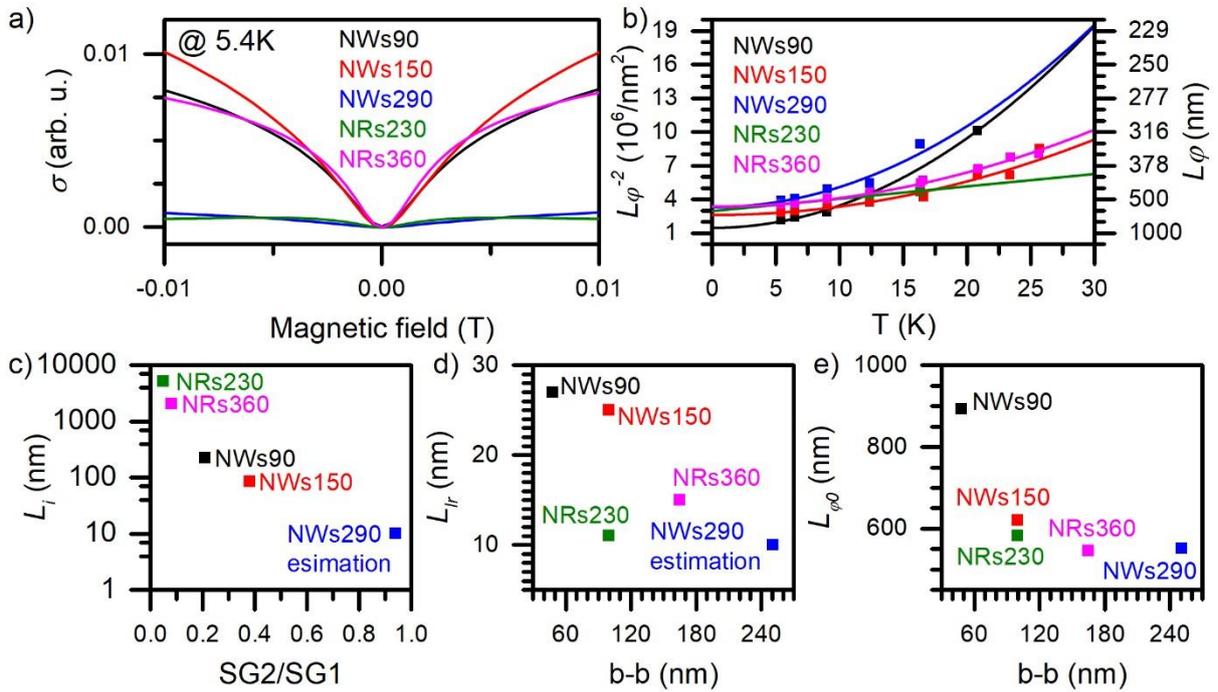

Fig. 6. Results of contactless transport measurements of weak localization in graphene on NWs and NRs: a) weak localization signal at 5.4 K, b) dependence of inelastic coherence scattering length on temperature, c) dependence of intervalley scattering length on SG2/SG1, d) dependence of long-range scattering length on b-b, e) dependence of coherence scattering length zero-offset on b-b.

Elastic intervalley scattering length reaches the largest value (more than several microns) in NRs230 and NRs360 (Tab. 3), while for NWs90 samples one order of magnitude lower value was obtained. For NWs150 its value further decreases 2.5 times while for NWs290 this scattering part could not be fitted (its value is lower than 10 nm). This phenomenon is not related to the creation of defects by deposition on supporting NWs/NRs due to the opposite correlation, however, correlation with strain gradient (SG2/SG1 ratio, Fig. 6c) is present. This strain gradient is a result of variations in distances between NWs/NRs borders. In NRs samples, most of the graphene surface is characterized by low variations of strain (and attributed to SG1 component in the deformation distribution). Only 4-5% of graphene located on dislocations in NRs arrangement is characterized by a significantly lower value of strain (SG2 component in the deformation distribution). In NWs samples, the percentage of points in SG2 component increases from 15% in NWs90 to 48% in NWs290; therefore, the ratio of SG2 to SG1 has an opposite trend than intervalley scattering length (Fig. 6c). This observation suggests that the strain gradient induces a local pseudomagnetic field which provides an efficient transfer of momentum.

A similar conclusion about the dependence between graphene strain and scattering process can be drawn from the analysis of the elastic long-range scattering length. This length is the lowest of all three scattering lengths (Fig. 6d). It has a similar value in NWs90, and NWs150 which is two times higher than for NRs samples. For NWs290 this scattering length was too small to extract its precise value, it was possible only to estimate its upper limit to be 10 nm. Thus, an inverse correlation of $L_{lr}$ length with distances between NWs/NRs borders (b-b) and the value of maximum graphene deformation is present (Fig. 6d). In NWs90 and NWs150 $L_{lr}$ is longer than in NRs samples. The presence of multiple wrinkles in NRs samples as well as a further increase of b-b could be responsible for the shortening of $L_{lr}$. Finally, the lowest intravalley scattering length is present in NWs290. This observation suggests the presence of an impact of a pseudomagnetic field on intravalley scattering mediated by wrinkles in graphene which confirms the Raman results.[12,66] Analysis of the temperature dependence of coherence scattering length shows that almost all samples show the electron-electron scattering in the ballistic regime (Fig. 6b). The only exception is NRs230 where the diffusive regime was observed. Some variations in the temperature dependence of coherence scattering length between samples could be an effect on the variations of temperature coefficients caused by differences in graphene Fermi level or electron diffusion coefficients.[55,56] For all the samples a zero-offset ($L_{\varphi 0}^{-2}$), was observed in $L_{\varphi}^{-2}(T)$ dependence. The presence of this offset can be attributed to the temperature-independent scattering mediated by an interaction with a pseudomagnetic field. Its value shows negative correlation with distances between NWs/NRs borders b-b (Fig. 6e).

Table 3, values of scattering lengths at 5.4 K: intervalley scattering lengths ($L_i$), long-range scattering lengths ($L_{lr}$) and value of coherence length zero-offsets ($L_{\varphi 0}$).

| Sample | $L_i$ (nm) | $L_{lr}$ (nm) | $L_{\varphi 0}$ (nm) |
|---|---|---|---|
| NWs90 | 220 | 27 | 893 |
| NWs150 | 85 | 25 | 620 |
| NWs290 | <10 | <10 | 551 |
| NRs230 | 5075 | 11 | 582 |
| NRs360 | 2035 | 15 | 546 |

In NRs samples $L_i$ length is significantly larger than $L_{lr}$ which shows that pseudomagnetic fields induced at graphene wrinkles provide a small transfer of momentum necessary for pseudospin flip and generation of intravalley scattering. A strain gradient that occurs at the borders between areas of graphene characterized by lower and higher deformation provides higher momentum which enables isospin flip and intervalley scattering. Therefore, our results suggest that the presence of graphene wrinkles is responsible for the occurrence of intravalley scattering due to the generation of pseudomagnetic field which is the dominant mechanism of scattering and that variations in b-b increase the intervalley scattering rate. The observed effect of weak localization suppression can be attributed to the induction of pseudomagnetic fields.[13]

The performed analysis of the weak localization effect and reduction of scattering lengths in graphene on NWs and NRs shows that the density of wrinkles presented in Fig. 2 is not the main factor responsible for the induction of pseudomagnetic fields. The coherence and intervalley scattering lengths in the NWs90 sample exhibiting a high density of wrinkles are significantly higher than in NWs290, where only a few wrinkles are present. Therefore, two parameters are important for the control of pseudomagnetic fields. Firstly, larger distances between borders of supporting regions increases graphene strain at wrinkles which is responsible for the induction of pseudomagnetic fields. Interestingly, multiple wrinkles effectively generate intravalley scattering but are less effective in the generation of intervalley scattering. Therefore, pseudomagnetic fields generated at wrinkles and local strain gradients provide lower or higher momentum for the occurrence of scattering, respectively. A local variation of distances between borders of supporting regions or the presence of NW/NR vacancy increases the local area of suspended graphene and strain gradient. This effect could be used for the fabrication of structures where pseudomagnetic field will be locally enhanced compared to graphene suspended between closely located supporting regions. Thus, the analysis of the WL signal and the estimated scattering lengths show the possibility of a wide control over the interaction with pseudomagnetic fields facilitated by the morphology of NWs/NRs substrate. Graphene deposited on NWs/NRs with various interdistances, and diameters can thus be interesting not only purely as a research object but also from the perspective of fabrication of pseudomagnetic field-based devices.

Conclusions

The presented study on graphene deposited on GaN NWs and GaN NRs highlights the potential of such structures for graphene strain engineering and the generation of pseudomagnetic fields.

Differences in NWs interdistances, diameters, and percentage of the supported graphene surface affect wrinkles formation. In all NWs samples, only the highest NWs are in contact with graphene. A similar behavior is present in NRs structures, which are characterized by more regular order than randomly distributed epitaxially grown NWs. Graphene wrinkles connect the nearest NWs and NRs while for NRs with larger diameters some of them are double or triple. Our analysis shows that HybriD mode AFM is a complementary technique for studies of strain-induced wrinkle formation, while SEM and tapping mode AFM are better for visualizing large graphene wrinkles. Moreover, statistical analysis of samples deformation enables to obtain more information than simple topography mapping. This approach enables to estimate the percentage of graphene which is supported by NWs/NRs, the strain distribution around them and the percentage of suspended graphene. Moreover, the strain gradient in graphene on NWs and NRs increases with increasing distances between NWs and NRs borders. AFM is supported by Raman spectroscopy which confirms the role of the percentage of graphene supported by NWs/NRs in the local reduction of graphene strain and its variations.

Graphene deposited on NWs or NRs is characterized by compressive strain while a tensile strain is present on graphene wrinkles. Suspended graphene areas reduce the value of strain in graphene. A lower uniformity of the arrangement of supporting regions also widens the spread of the strain values. An analysis of the D' band intensity suggests the presence of an intravalley scattering enhances by pseudomagnetic fields. Our study of the local density of states performed on suspended graphene wrinkle show modulation of the electronic properties in both directions i.e., perpendicular and along the wrinkle. Formation of the additional peak in the vicinity to the Fermi level and the energy shift of Dirac point location were observed. The additional spectral feature at the Fermi level could be a fingerprint of the pseudo-landau levels. Finally, our analysis of the weak localization signal shows the presence of a correlation between an increase of the distances between NWs/NRs borders and a gradient of strain in graphene with the reduction of long-range, intervalley scattering length, and coherence scattering lengths. These results confirm that an interaction with a pseudomagnetic field in graphene on NWs and NRs occurs on a large area, while the strength of an interaction depends on the distances between supporting regions borders and their arrangement. Thus, our results show a good perspective for controlling the local gradient of strain in graphene supported on NWs or NRs providing the possibility of the induction of pseudomagnetic fields. A wide control of the graphene morphology and spatial distribution of graphene wrinkles is also promising for future studies of these structures and their applications for sensors or valley filters.

Methods

GaN NWs were grown on *in-situ* nitridated Si(111) substrate in a Riber Compact 21 system with an elemental source of Ga and Al. A radio frequency Addon nitrogen plasma cell, controlled by an optical sensor of plasma light emission was used as the source of active nitrogen species.[67] The growth temperature and time were varied to prepare NWs with different: interdistances and number of NWs on the surface. Firstly, silicon substrate was nitridated by 30 minutes in 850°C in active nitrogen flux of 6 nm/min.[30] Then, substrate was cooled down to 760°C and Ga source was opened. NWs90 sample was grown for 120 minutes under 3.9 nm/min Ga flux. In NWs150 and NWs, the Ga flux was 0.6 nm/min, while growth times were 250 min. and 540 min., respectively. Ga and N fluxes used for the growth were calibrated in GaN equivalent growth rate units (nm/min).[68]

To fabricate NRs, self-assembled colloidal monolayers of silica nanospheres of the desired diameter were scooped from the surface of the water on the surface of epitaxially growth Ga-polar GaN/sapphire templates. Using the silica nanospheres as a hard mask, a dry etching by chlorine-based inductively coupled plasma was used for the NRs formation. Further details on the GaN NR fabrication can be found elsewhere.[39] Residuals of silica nanospheres were then removed in HF acid solution.

Graphene transferred onto a solid epitaxial GaN/sapphire template (the same as the one used for NR fabrication) was used as the reference sample for Raman analysis and named GaNN.[69] Graphene was grown on a copper foil by chemical vapor deposition using methane as a carbon precursor.[70] Monolayer graphene is usually too fragile to be transferred onto corrugated substrates without cracking.[28] To prepare samples with large areas of high-quality transferred graphene we used two-layer graphene. It was obtained by transferring one graphene monolayer grown on a copper substrate onto another graphene monolayer by high-speed electrochemical delamination method.[70]

Raman spectra were measured with a Renishaw InVia spectrometer equipped with x100 objective under 532 nm laser excitation. Raman micromapping was performed on an area of several square micrometers with 100 nm step resulting in maps each of which consisting of 441 points. The laser power was kept at several mW to minimize the heating effects.

Scanning electron microscopy (SEM) measurements were performed with a Helios Nanolab 600 dual beam system. A secondary electron detector at 5 kV electron beam voltage and 86 pA current was used. AFM measurements were performed using NTEGRA Aura (NT-MDT) in semicontact mode and in HybriD mode, where force-versus-distance curves were recorded in each raster point using 1.5 kHz sine vibration frequency with a typical amplitude of 30 nm applied to the sample (allowing to record images of sample height, adhesion, deformation, and stiffness).

STM was performed with a mechanically cut tip of 90% Pt-10% Ir alloy wires (Goodfellow) with a VT-STM/AFM microscope (Scienta-Omicron) operating in UHV at $10^{-8}$ Pa. STM scans were acquired at a bias of - 3 V. In the current imaging tunneling spectroscopy mode (CITS) the I/V curves were recorded simultaneously with topography images by the interrupted-feedback-loop technique and then numerical first derivative and normalization calculated (by dividing dI/dV(V) by I/V). Contactless weak localization measurements were based on the measurements of changes in the microwave-cavity Q factor (quality factor) with the applied magnetic field in an Electron Spin Resonance (ESR) Bruker ELEXSYS E580 spectrometer.[36] The spectrometer operates at a microwave frequency of 9.4 GHz (X-band) with a $TE_{102}$ resonance cavity. It is equipped with a helium cryostat, allowing to cool the sample down to 5.4 K. During the measurements, the microwave power and the modulation amplitude were set to 0.47 mW and to 0.1 mT, respectively.

Acknowledgments

This work was partially supported by the Ministry of Science and Higher Education in years 2015-2019 as a research grant "Diamond Grant" (No. DI2014 015744), National Science Centre, Poland, under the Grant Nos. 2018/31/B/ST3/02450 (PJK, MR and PK), 2018/30/E/ST5/00667 (PD) and 2016/23/B/ST7/03745 (MS and ZRZ).

We would like to extend our most sincere acknowledgments to our late colleagues: dr Krzysztof Pakuła, Kamil Klosek and prof. Zbigniew Klusek for their support with sample preparation, conducting scientific research, and data analysis.